\documentclass{article}

\usepackage[english]{babel}

\usepackage[letterpaper,top=2cm,bottom=2cm,left=3cm,right=3cm,marginparwidth=1.75cm]{geometry}

\usepackage{amsmath}
\usepackage{graphicx}
\usepackage[colorlinks=true, allcolors=blue]{hyperref}
\usepackage{multirow}
\usepackage{setspace}
\usepackage{authblk}

\title{Multiple Imputation Methods for Missing Multilevel Ordinal Outcomes}
\author[1]{Mei Dong}
\author[1*]{Aya Mitani}
\affil[1]{Department of Biostatistics, Dalla Lana School of Public Health, University of Toronto, Ontario, Canada}

\date{}

\begin{document}
\maketitle

\begin{abstract}

Multiple imputation (MI) is an established technique to handle missing data in observational studies. Joint modeling (JM) and fully conditional specification (FCS) are commonly used methods for imputing multilevel clustered data. However, MI approaches for ordinal clustered outcome variables have not been well studied, especially when there is informative cluster size (ICS). The purpose of this study is to describe different imputation and analysis strategies for the multilevel ordinal outcome when ICS exists. We conducted comprehensive Monte Carlo simulation studies to compare five different methods: complete case analysis (CCA), FCS, FCS+CS (include cluster size (CS) when performing the imputation), JM, and JM+CS under different scenarios. We evaluated their performances using an proportional odds logistic regression model estimated with cluster weighted generalized estimating equations (CWGEE). The simulation results show that including cluster size in imputation can significantly improve imputation accuracy when ICS exists. FCS provides more accurate and robust estimation than JM, followed by CCA for multilevel ordinal outcomes. We further applied those methods to a real dental study.

\end{abstract}

\section{Introduction}\label{sec1}
Multilevel ordinal outcomes commonly appear in observational studies. In a study of dental disease, maximum clinical attachment loss (CAL) -- recorded on each tooth using an ordinal scoring system -- was used to assess periodontal health. In this context, each subject contributes multiple outcomes of interest (CAL) and they are clustered within a subject. Therefore, the correlation between the outcomes within the same subject need to be accounted for in the analysis. Generalized linear mixed effect model (GLMM) and generalized estimating equation (GEE), both extensions of the generalized linear model (GLM), are the two most popular methods to model such multilevel data. GLMM gives cluster-specific inference while marginal models using GEE give population-average inference. 

Both GLMM and GEE assume that cluster size and the outcome of interest are independent, given the covariates. However, this assumption can be violated when cluster size varies and changes with the degree of the outcome. For example, in periodontitis, the probability of losing a tooth increases with the severity of the disease. Hence, patients with advanced periodontitis tend to have fewer number of teeth, or smaller cluster size, compared to patients with good oral health. This type of situation, when the outcome is dependent on cluster size conditional on observed covariates, is referred to as informative cluster size (ICS). To handle ICS, Seaman et al. summarized a number of methods based on GLMM and GEE \cite{seaman2014methods}. Including the cluster size as a covariate in the model is one of the solutions, but it changes the interpretations of coefficients of other covariates. When the interest is in making marginal inference, Williamson et al. and Benhin et al. both proposed cluster weighted GEE (CWGEE), which provides unbiased estimator when ICS exists \cite{williamson2003marginal, benhin2005cwgee} and can model ordinal outcomes \cite{mitani2019marginal}.  

While CAL cannot be measured on missing teeth (which leads to the issue of ICS), in the dental study, some CAL measurements were missing also on existing teeth for unknown reasons. Removing the teeth with missing CAL values from the analysis will produce inconsistency between cluster size and the number of observed outcomes for some of the subjects. Furthermore, marginal models using GEE assume that missing data are unrelated to observed and unobserved variables. Therefore, when the interest is in making marginal inference, missing data problems are often dealt by multiple imputation (MI) \cite{schafer1997analysis, little2019statistical}. Briefly, MI  involves three phases: imputation phase, analysis phase, and pooling phase. In the imputation phase, the missing values are filled in with plausible values estimated from the posterior predictive model. This process is repeated $M$ times, creating $M$ different complete datasets. Then in the analysis phase, statistical analysis is performed on each of the $M$ complete datasets, leading to $M$ estimates and variance-covariance matrices of the parameters. Finally, in the pooling phase, the $M$ parameter estimates and variances are pooled to create one set of parameter and variance estimates \cite{rubin2004multiple}. 

There are two main strategies for MI: joint modeling (JM) and fully conditional specification (FCS), also known as MI with chained equation (MICE). Schafer proposed a joint multivariate normal model, which assumes that partially observed data follow a multivariate normal distribution \cite{schafer1997analysis}. FCS is known for its flexibility in handling data with mixed variable types by imputing the variables containing missing values one at a time. Hence, FCS requires a unique specification of the imputation model for each variable with missing values \cite{vanBuuren2018book}. Both strategies have been implemented in mainstream statistical programming languages, including R, and standalone software.  R packages that perform MI based on JM include \textbf{norm} \cite{novo2002analysis}, 
\textbf{cat} \cite{harding6cat}, 
\textbf{mix} \cite{schafermix},  
\textbf{pan} \cite{pan} and \textbf{jomo} \cite{quartagno2019jomo}. There is also a standalone software \textbf{REALCOM} \cite{carpenter2011realcom} that performs JM imputation. However, only \textbf{REALCOM} and \textbf{jomo} are able to impute multilevel categorical data. The R package \textbf{mice} is the most commonly used R package to implement FCS, which provides many options for model specification \cite{van2011mice}. However, \textbf{mice} has limited options to impute multilevel data. Other  packages, such as \textbf{micemd} \cite{audigier2019micemd} and \textbf{miceadds} \cite{robitzsch2017package}, as extensions for \textbf{mice}, provide more options for different types of variables in multilevel data.  Nevertheless, \textbf{micemd} does not deal with ordinal data and \textbf{miceadds} uses predictive mean matching to impute ordinal data. Recently, Enders et al. developed a standalone software \textbf{blimp} that uses a latent probit approach to impute multilevel ordinal data \cite{enders2018fully}. 

Currently, the R package \textbf{jomo} and the standalone software \textbf{blimp} are two optimal software for imputing missing multilevel ordinal data. Although their performances on imputing multilevel continuous and binary data have been compared in many different aspects \cite{enders2016multilevel,audigier2018multiple,  wijesuriya2022multiple}, their performances on imputing multilevel ordinal outcomes, especially when ICS exists, have not been studied yet. Kombo et al. compared FCS and JM for ordinal longitudinal data with monotone missing data patterns \cite{kombo2017multiple}, but many multilevel data, such as clustered dental data, do not follow a monotone missing data pattern. 

The objective of this paper is to compare the performances between JM and FCS when imputing clustered ordinal outcomes that are subject to ICS. Two available software packages will be used: \textbf{jomo} in R for JM and \textbf{blimp} for FCS. For each of the JM and FCS approaches, we will additionally assess whether the inclusion of cluster size in the imputation model will improve parameter estimation of the analysis model. Since we are interested in the population-average inference, parameters in the analysis model will be estimated by CWGEE with proportional odds logit. The rest of the paper is organized as follows. In section \ref{sec2}, we describe the methods used for the analysis model and imputation models. Extensive simulation studies as well as simulation results are shown in section \ref{sec3}. In section \ref{sec4}, we apply the imputation methods to a real dental data. Finally,  conclusions and discussion are given in section \ref{sec5}.

\section{Methods}\label{sec2}

\subsection{Motivating example: the VADLS study}
Our study is motivated by the Veterans Affairs Dental Longitudinal Study (VADLS), which was a closed-panel longitudinal study that monitored oral health and diseases of male subjects from the greater Boston metropolitan area  \cite{kapur1972veterans}. The health status of the subjects were measured approximately every three years. For illustration, we focused on one cycle of the longitudinal data, which includes 241 subjects with a total of 5100 teeth. We were interested in assessing the association between metabolic syndrome (MetS) and increasing CAL scores \cite{kaye2016metabolic}. CAL score is a level-1 (tooth/member level) ordinal variable of four categories (0: $<$ 2mm, 1: 2-2.9mm, 2: 3-4.9mm, 3: $\ge$ 5mm). The higher score indicates a worse prognosis of periodontal disease. We modelled the association between MetS (yes/no) and ordinal CAL scores using the proportional odds logistic regression adjusted for level-2 (subject/cluster level) variables: age, smoking status (ever-smoker/never-smoker), and education levels (high school, some college, college graduate). The summary statistics and missing percentages of all variable are shown in Table \ref{tab:char}. These variables have been shown to be associated with CAL from previous studies \cite{kaye2016metabolic, gamonal2010clinical}. The marginal analysis model had the following form:
\begin{equation}
\begin{aligned}\label{CAL}
           \text{logit}\left\{\Pr(\text{CAL}_{ij} \leq c)\right\} =& \eta_c + \beta_1 \text{MetS}_i +\beta_2 \text{age}_i + \beta_3 \text{smoking}_i   \\
           & + \beta_4 1(\text{edu}_i=\text{some college})+\beta_5 1(\text{edu}_i=\text{college graduate}), c=1,2,3 
\end{aligned}
\end{equation}
where $\text{CAL}_{ij}$ is CAL recorded on the $j$th tooth of the $i$th subject, $j=1,...,n_i$, $i=1,...,N$. Two issues existed in producing valid inference from Equation (\eqref{CAL}). First, the number of teeth per subject ranged from 1 to 28 and the Spearman correlation coefficient of mean CAL score per subject and the number of teeth per subject was -0.41 (95\% CI: -0.44, -0.38), indicating the presence of ICS in this data. Therefore, CWGEE was applied for estimation. Second, CAL was missing in 19\% of all existing teeth. Hence, MI was applied to make use of all available data before model fitting. In addition to CAL, three other level-1 variables that measure the prognosis of periodontitis were available: probing pocket depth (PPD), alveolar bone loss (ABL) and mobility (Mobil). PPD, ABL, and Mobil were also recorded using an ordinal scoring system and were correlated with each other and with CAL. Although PPD, ABL, and Mobil were not included in the analysis model, they were included in the imputation model to help impute missing CAL.

\subsection{Ordinal regression with CWGEE}

In the dental study, our goal was to obtain the marginal inference of association between MetS and the periodontal health of a typical tooth from a randomly selected person. GEE is appealing not only because the estimator of coefficient is almost as efficient as the maximum likelihood estimator but also because it provides a consistent estimator of coefficient even under a misspecified within-subject association among the repeated measurements when we have sufficiently large samples \cite{fitzmaurice2008longitudinal}. Due to ICS, estimation using GEE will have oversampled healthy teeth and resulted in biased coefficients \cite{seaman2014review}. To overcome this challenge, Williamson et al. and Benhin et al. both proposed CWGEE \cite{williamson2003marginal, benhin2005cwgee}, which weighs the GEE by the inverse of cluster size. 
Let $Y_{ij}$ denote the ordinal outcome $C>2$ categories for the $j$th member of the $i$th subject, $i=1,..., N$, $j=1, .., n_i$, where $n_i$ is the cluster size (number of teeth) for subject $i$. Let $\boldsymbol{X}_{ij}=(X_{ij1},...,X_{ijp})^T$ denote the sets of $p$ fixed covariates for the $j$th member of the $i$th subject. The model for ordinal outcome using proportional odds logistic regression is expressed as
\begin{equation*}
    \text{logit}\{\Pr(Y_{ij} \leq c)\}=\eta_c+\boldsymbol{X}_{ij}^T \boldsymbol{\beta}, \quad c=1,...,C-1.
\end{equation*}
For estimation, it is common to re-express the $C$-category outcome $Y_{ij}$ as a set of $C-1$ binary outcomes, such that
$$
U_{ijc} = 
\begin{cases}
1 & Y_{ij} \leq c, \\
0 & Y_{ij} > c
\end{cases} \quad c=1,...,C-1
$$
and write the model as a set of logistic regression for each of the $C-1$ binary outcomes \cite{kenward1994ordinal}:
\begin{equation*}
    \text{logit}\{\text{E}(U_{ijc})\}=\text{logit}\{\Pr(U_{ijc} = 1)\}=\eta_c+\boldsymbol{X}_{ij}^T \boldsymbol{\beta}, \quad c=1,...,C-1.
\end{equation*}
Let $\boldsymbol{\mu_{ij}}=\text{E}(\boldsymbol{U}_{ij})$, where $\boldsymbol{U}^{T}_{ij} = (U_{ij1}, ..., U_{ijC-1})$. Then the estimation of $(\eta_1, ..., \eta_{C-1}, \boldsymbol{\beta})$ is obtained by solving the following CWGEE,
\begin{equation}\label{CWGEE}
\sum_{i=1}^N \frac{1}{n_i}\sum_{j=1}^{n_i}\boldsymbol{D}_{ij}'\boldsymbol{V}_{ij}^{-1}(\boldsymbol{U}_{ij}-\boldsymbol{\mu}_{ij})=0,
\end{equation}
where $\boldsymbol{D}_{ij}=\partial \boldsymbol{\mu}_{ij}/{\partial  \boldsymbol{\beta}}$, $\boldsymbol{V}_{ij}=\boldsymbol{A}_{ij}^{1/2} \boldsymbol{R}_{ij} \boldsymbol{A}_{ij}^{1/2}$ and $\boldsymbol{A}_{ij}$ is the diagonal matrix of variance $\text{Var}(\boldsymbol{\mu}_{ij})$ and $R_i$ includes the pairwise correlation between $U_{ijc}$ and $U_{ijc'}$ for $c \neq c', c=1,...,C-1$. Note that with this estimating equation, we assume a ``working independence" structure between the teeth within a subject which is conventional in CWGEE \cite{Williamson07}. A robust standard error (se) is  estimated using the sandwich variance estimator, which has the form
\begin{equation*}
   \hat{\Psi}=\hat{H}^{-1} \hat{M} \hat{H}^{-1},
\end{equation*}
where
\begin{equation*}
\hat{H}=\sum_{i=1}^N \frac{1}{n_i}\sum_{j=1}^{n_i}\hat{D}'\hat{\boldsymbol{V}}_{ij}^{-1} \hat{D},
\end{equation*}
and 
\begin{equation*}
\hat{M}=\sum_{i=1}^N [\frac{1}{n_i}\sum_{j=1}^{n_i}\hat{D}'\hat{\boldsymbol{V}}_{ij}^{-1}(\boldsymbol{U}_{ij}-\boldsymbol{\mu}_{ij})] [\frac{1}{n_i}\sum_{j=1}^{n_i}\hat{D}'\hat{\boldsymbol{V}}_{ij}^{-1}(\boldsymbol{U}_{ij}-\boldsymbol{\mu}_{ij})]'.
\end{equation*}

\subsection{MI with multilevel ordinal data}
MI is a Monte Carlo technique in which missing values are replaced by a set of simulated values drawn from the posterior predictive distribution $P(Y_{miss}|Y_{obs})$. JM and FCS are two main strategies for MI: JM assumes a multivariate normal model for all variables, while FCS specifies a single model for each variable and imputes each of them sequentially. For either strategy, the imputation phase can be summarized into two steps:
\begin{enumerate}
    \item Draw $\boldsymbol{\theta}$, the parameters of the imputation model, from the posterior distribution using the compete data;
    \item Update the imputation by drawing data from the posterior predictive model $P(Y_{miss}|Y_{obs}, \boldsymbol{\theta})$.
\end{enumerate}
The first step utilizes Gibbs sampler, which is iterated many times to yield an empirical estimate of each parameter's marginal posterior distribution \cite{enders2018fully}. Each approach has been extended to handle imputation of multilevel data by including a random effects term in the imputation model to account for the correlation between members within the same cluster \cite{enders2016multilevel, enders2018fully}.

Following the variables in the dental study, we denote $Y_{ij}$ (CAL) as the level-1 ordinal outcome with missing data with $C_Y$ categories. We also have three ``auxiliary" level-1 ordinal variables, $M_{1ij}$ (PPD), $M_{2ij}$ (ABL), and $M_{3ij}$ (Mobil) with $C_{M_1}$, $C_{M_2}$, and $C_{M_3}$ categories respectively. The ``auxiliary" variables are not of direct interest in the analysis but can improve imputation accuracy if included in the imputation model. In addition, we have a level-2 continuous covariate $X_{i}$ and a level-2 binary covariate $Z_{i}$, which are both fully observed. These covariates are in both the analysis and imputation models.

\subsubsection{Joint Modeling with R package jomo}
JM models the data through a multivariate normal distribution. In the R package \textbf{jomo}, categorical variables are substituted with latent normal variables during Gibbs sampling and then converted back to discrete values using thresholds \cite{quartagno2019jomo}. For categorical variables with $C$ levels, we need $C-1$ latent normal variables, each of which has a fixed variance of 1, and the covariance with the other latent normal variables of 0.5 \cite{quartagno2019multiple}. To deal with multilevel data, multivariate version of LME is  used. In this paper, we consider the random intercept model since only the level-1 outcomes contain missing data. Let $Y_{ijc}^*$, $M_{1ijc}^{*}, M_{2ijc}^{*}$, and $M_{3ijc}^{*}$ be the latent normal variables for $Y_{ij},M_{1ij}, M_{2ij}, M_{3ij}$ in level $c$, respectively. Then, we construct a multivariate random intercept model as the multilevel imputation model:
\begin{align*} 
Y_{ijc}^* &= \beta_{0c,Y}+\gamma_{1,Y}X_i + \gamma_{2,Y}Z_i+u_{Y,i,c}+\epsilon_{Y,ij,c}, c=1,...,C_Y-1 \\ 
M_{1ijc}^{*} &= \beta_{0,M_1}+\gamma_{1,M_1}X_i + \gamma_{2,M_1}Z_i+u_{M_1,i,c}+\epsilon_{M_1,ij,c}, c=1,...,C_{M_1}-1 \\
M_{2ijc}^{*} &= \beta_{0,M_2}+\gamma_{1,M_2}X_i + \gamma_{2,M_2}Z_i+u_{M_2,i,c}+\epsilon_{M_2,ij,c}, c=1,...,C_{M_2}-1 \\
M_{3ijc}^{*} &= \beta_{0,M_3}+\gamma_{1,M_3}X_i + \gamma_{2,M_3}Z_i+u_{M_3,i,c}+\epsilon_{M_3,ij,c}, c=1,...,C_{M_3}-1 
\end{align*}
\begin{equation*}
\text{with }
\begin{pmatrix}
u_{Y,i, 1}\\
\dots \\
u_{Y,i, C_Y-1} \\
u_{M_1,i,1} \\
\dots \\
u_{M_3,i, C_{M_3}-1}
\end{pmatrix} \sim MVN(\boldsymbol{0}, \boldsymbol{\Sigma}_u), \quad
\begin{pmatrix}
\epsilon_{Y,ij,1}\\
\dots \\
\epsilon_{Y,ij,C_Y-1}\\
\epsilon_{M_1,ij,1} \\
\dots  \\
\epsilon_{M_3,ij, C_{M_3}-1}
\end{pmatrix} \sim MVN(\boldsymbol{0}, \boldsymbol{\Sigma}_\epsilon)
\end{equation*}
Then, the imputed latent variables were converted back to discrete values as follows: 
\begin{equation*}
    Y_{ij}=\begin{cases}
      1, &  Y_{ij1}^* > 0  \hspace{0.5em} \& \hspace{0.5em} Y_{ij1}^* > Y_{ijc}^* \hspace{0.5em} \text{for} \hspace{0.5em} c \neq 1, C_Y\\
      2, & Y_{ij2}^* > 0 \hspace{0.5em} \& \hspace{0.5em} Y_{ij2}^* > Y_{ijc}^* \hspace{0.5em} \text{for} \hspace{0.5em} c\neq 2, C_Y \\
      \dots \\
      C, & Y_{ijc}^* < 0 \hspace{0.5em} \text{for} \hspace{0.5em} c \neq C_Y
    \end{cases}   
\end{equation*}
and similarly for $M_{1ij}, M_{2ij}, M_{3ij}$.

\subsubsection{Fully Conditional Specification with blimp software}
Rather than assuming the variables with missing data to follow a multivariate normal distribution, FCS assumes a unique model for each variable with missing values. Based on the distribution of the variable, a unique type of imputation model can be specified (e.g. linear regression for continuous variable, logistic regression for binary variable). For multilevel data, an additional step is required for the Gibbs sampling steps, which is generating imputations based on the current model parameters and level-2 residual terms. Thus, each Gibbs cycle uses the current imputations to execute a complete-data Bayesian analysis \cite{enders2018fully}. The procedure of multilevel FCS can be summarized as follows for iteration $t$:
\begin{enumerate}
    \item $Y_{ij}$ is drawn from the following distribution:
    \begin{align*}
   &  Y_{ij(miss)}^{(t)} \sim N(\hat{Y}_{ij}^{(t)}, \sigma_{\epsilon(Y)}^2) 
    \\
   &  \hat{Y}_{ij}^{(t)}  =\beta_{0Y}+\beta_{1Y}M_{1ij}^{(t-1)} +\beta_{2Y}M_{2ij}^{(t-1)} +\beta_{3Y}M_{3ij}^{(t-1)}+\gamma_{1Y}X_{i}+\gamma_{2Y}Z_{i}+u_{0Y},
    \end{align*} 
    where $\hat{Y}_{ij}^{(t)}$ is the predicted value from the multilevel model and $\sigma_{\epsilon(Y)}^2$ is the within-cluster residual vairance.
    
    \item After updating $Y_{ij}$, $M_{1ij}$ is now treated as the outcome and $Y_{ij}$ and other covariates are treated as predictors. $M_{1ij}$ is drawn from the following distribution:
    \begin{align*}
       &   M_{1ij(miss)}^{(t)} \sim N(\hat{M}_{1ij}^{(t)}, \sigma_{\epsilon(M_1)}^2) \\
       &   \hat{M}_{1ij}^{(t)}=\beta_{0M_1}+\beta_{1M_1}Y_{ij}^{(t-1)} +\beta_{2M_1}M_{2ij}^{(t-1)} +\beta_{3M_1}M_{3ij}^{(t-1)}+\gamma_{1M_1}X_{i}+\gamma_{2M_1}Z_{i}+u_{0M_1}
    \end{align*}
    \item After updating $M_{1ij}$, repeat the above steps to update $M_{2ij}$ and $M_{3ij}$.
\end{enumerate}

The described steps are standard algorithm for multilevel FCS, which are implemented in the R package \textbf{mice} \cite{van2011mice}. This approach is flexible and useful in many applications. However, the imputation model options in \textbf{mice} are limited to incomplete level-1 variables that are either continuous or binary. Enders et al. extended \textbf{mice} to handle incomplete nominal and ordinal variables, and provided the software program \textbf{blimp} \cite{enders2018fully}. 

In \textbf{blimp}, for ordinal data, the cumulative probit model is used to link the latent variable distribution to the discrete responses using a threshold parameter. The imputation model for $Y_{ij}$ in step 1 then becomes:
  \begin{align} \label{FCS}
   &  Y_{ij(miss)}^{*(t)} \sim N(\hat{Y}_{ij}^{*(t)}, \sigma_{\epsilon(Y)}^2) \nonumber
    \\
   &      Y_{ij}^{*(t)}=\beta_{0Y^*}+\beta_{1Y^*}M_{1ij}^{*(t-1)} +\beta_{2Y^*}M_{2ij}^{*(t-1)} +\beta_{3Y^*}M_{3ij}^{*(t-1)}+\gamma_{1Y^*}X_{i}+\gamma_{2Y^*}Z_{i}+u_{iY^*}.
    \end{align}
Then the imputed latent variable is converted to ordinal variable using the following function: \begin{equation*}
    Y_{ij}=f(Y_{ij}^*)=\begin{cases}
      1, & -\infty <Y_{ij}^*<\tau_1 \\
      2, & \tau_1 <Y_{ij}^*<\tau_2 \\
      \dots \\
      C, & \tau_{C-1} <Y_{ij}^*<\infty,
    \end{cases}   
\end{equation*}
which finalize the imputation of $Y$ in iteration $t$. We then repeat step 2 and step 3 similarly to other variables $M_{1ij}$, $M_{2ij}$, and $M_{3ij}$.

\subsubsection{Imputation model for data with ICS}

In MI, the imputation model needs to include all the features of the analytical model \cite{sterne2009multiple}. The existence of ICS indicates that the outcome $Y_{ij}$ is dependent on the cluster size $n_i$ and ignoring this relationship in the imputation model may lead to inefficient and biased estimation of the posterior distribution. Hence, we will include $n_i$ in the imputation model to account for the dependence between the missing $Y_{ij}$ values and $n_i$. Taking FCS as an example, Equation (\ref{FCS}) is then rewritten as 
\begin{equation*}\label{FCS2}
    Y_{ij}^{*(t)}=\beta_{0Y^*}+\beta_{1Y^*}M_{1ij}^{*(t-1)} +\beta_{2Y^*}M_{2ij}^{*(t-1)} +\beta_{3Y^*}M_{3ij}^{*(t-1)}+\gamma_{1Y^*}X_{i}+\gamma_{2Y^*}Z_{i}+\gamma_{3Y^*}*n_{i}+u_{iY^*}.
  \end{equation*} 
Note that $n_i$ will only be included in the imputation phase to account for additional variance of $Y_{ij}$, thus potentially improving the imputation accuracy. In the analysis model, ICS is accounted for by the CWGEE as in Equation (\ref{CWGEE}). In a simulation study, we will compare the effect of including versus omitting $n_i$ in the imputation model for both JM and FCS. 

\section{Simulation Studies}\label{sec3}

\subsection{Simulation Setting}
We assessed the performance of five missing data approaches for multilevel ordinal outcomes with ICS through a comprehensive simulation study: 1) Complete-case analysis (CCA); 2) FCS without cluster size as a predictor (FCS); 3) FCS with cluster size as a predictor (FCS+CS); 4) Joint modeling without cluster size as a predictor (JOMO); 5) Joint modeling with cluster size as a predictor (JOMO+CS).

We mimicked the real dental data to generate the multilevel ordinal outcomes of $C=4$ categories. To generate clustered ordinal data that is subject to ICS, we simulated the outcome $Y_{ij}$ of the $j$th tooth of the $i$th subject from a proportional odds logit model with a random intercept that follows a bridge distribution:
\begin{equation}
    \text{logit}(\text{Pr}(Y_{ij} \le c))=\eta_c+\beta_1x_i+\beta_2z_i, \quad c=1,2,3, \quad j=1...,n_i, \quad i = 1,...,N
\end{equation}
where $x_i$ was a level-2 continuous variable generated from $\text{N}(0, 2^2)$, $z_i$ was a level-2 binary variable generated from $\text{Binomial}(N, 0.5)$. The true values for the parameters were $(\eta_1, \eta_2, \eta_3,\beta_1, \beta_2 )=(-0.4, 0.8, 1.6, -0.2, -0.5)$. We also simulated three other auxiliary level-1 ordinal variables $M_{1}, M_{2}$, and $M_{3}$ following the same procedure with different values of intercepts $(\eta_1, \eta_2, \eta_3)$. $M_{1}, M_{2}$, and $M_{3}$ were used in the imputation of $Y$.

To simulate the ordinal outcome with ICS, we used the bridge distribution \cite{mitani2019marginal, parzen2011generalized} to obtain the marginal probability of success when fitting a random intercept logistic regression model of the form:
\begin{equation}
    p_{ijc}=\text{Pr}(Q_{ijc}=1|b_i, x_{ij}, \boldsymbol{\beta})=\frac{\exp(b_i +\phi^{-1}x_{ij}' \boldsymbol{\beta})}{1+\exp(b_i +\phi^{-1}x_{ij}' \boldsymbol{\beta})},
\end{equation}
where $Q_{ijc}=1$ means the $j$th tooth for $i$th subject is in category $c$, and $b_i$ acts as a random intercept that follows a bridge distribution with density $f_b(b_i | \phi)=\frac{1}{2 \pi}\frac{\sin(\phi \pi)}{\cos(\phi b_i) +\cos(\phi \pi)}$, $-\infty < b_i < \infty$, $0 < \phi <1$.
The maximum cluster size in each subject is 28. We used the exchangeable correlation structure with parameter $\tau$ to simulate the correlation between teeth. For each subject $i$, we generated the baseline hazard $\lambda_i$ as a function of the subject specific set of random effects $b_i$, $\lambda_i=\frac{\exp(\nu \bar{b}_i))}{1+\exp(\nu \bar{b}_i))}$, where $\bar{b}_i=\sum_j \frac{b_{ij}}{n_i}$, and $\nu$ controls the degree of ICS. The number of tooth in the subjects was generated from $\text{Binomial}(28, \lambda_i)$. A detailed description of the simulation is shown in the Supplementary Materials.

We generated missing values through three different missing data mechanisms, MCAR, missing at random (MAR), missing not at random (MNAR) \cite{rubin1976inference} and considered different levels of missingness on the outcome $Y$. To generate the missing data mechanism, the missingness indicators $R_{ij}$ for the $j$th tooth of the $i$th subject were generated from the model:
\begin{equation}
    \text{logit}\{P(R_{ij}=1)\}=\alpha_0+\alpha_1 x_{i} + \alpha_2 z_{i} + \alpha_3 y_{ij} + \alpha_4 M_{1ij} +\alpha_5 M_{2ij} + \alpha_6 M_{3ij}
\end{equation}
When $\alpha_1=\dots =\alpha_6=0$, data were MCAR. When only $\alpha_3=0$, data were MAR. Otherwise, data were MNAR. $\alpha_0$ is used to control the missing rate. For the outcome $Y$, we generated missing rates of 20\% and 50\%, representing low and high missing rates, respectively. For the auxiliary outcomes $M_{1}, M_{2}$, and $M_{3}$, the missing rates were 30\%, 30\%, 10\%, respectively.

Table \ref{tab:para} shows the combination of the various parameters when data were MAR. We considered two different sample sizes $N$, 50 and 250. The degrees of correlation $\tau$ varied from 0, 0.3, to 0.6, representing null, moderate, and strong correlation between teeth. We varied the degrees of ICS $\nu$ from 0, 0.1, to 0.4, representing null, moderate, and high correlation between the outcome and cluster size. When data were MCAR and MNAR, we considered the scenario where $N=50, \tau=0.3, \nu=0.1$ and the missing rate was 20\%. We obtained the parameter estimates $(\hat{\eta}_1, \hat{\eta}_2, \hat{\eta}_3, \hat{\beta}_1, \hat{\beta}_2)$ from each simulated data in each scenario through 1000 replications. The following metrics were computed to compare the performance of each imputation approach: (1) the mean parameter estimates (Mean Est); (2)the mean robust standard error (Mean SE); (3) the empirical standard error (Empirical SE); (4) the mean relative bias * 100\% (Rel Bias); (5) the 95\% coverage probability (Cov Prob); and (6) the mean square error (MSE).

\begin{table}[]
    \centering
    \begin{tabular}{c|c|ccc}
    \hline
    Parameter & Notation &  & Values &  \\
    \hline
        Sample size & N & 50 & 250 \\
        Degree of correlation (ICC) & $\tau$ & 0 & 0.3 & 0.6 \\
        Degree of ICS & $\nu$ & 0 & 0.1 & 0.4 \\
        Missing rate & r & 20\% & 50\% & \\
        \hline
    \end{tabular}
    \caption{Parameter setings in the simulation study when data were MAR.}
    \label{tab:para}
\end{table}

\subsection{Results}
\begin{figure}[!htb]
    \centering
    \includegraphics[scale=0.6, trim={0 2cm 0 0},clip]{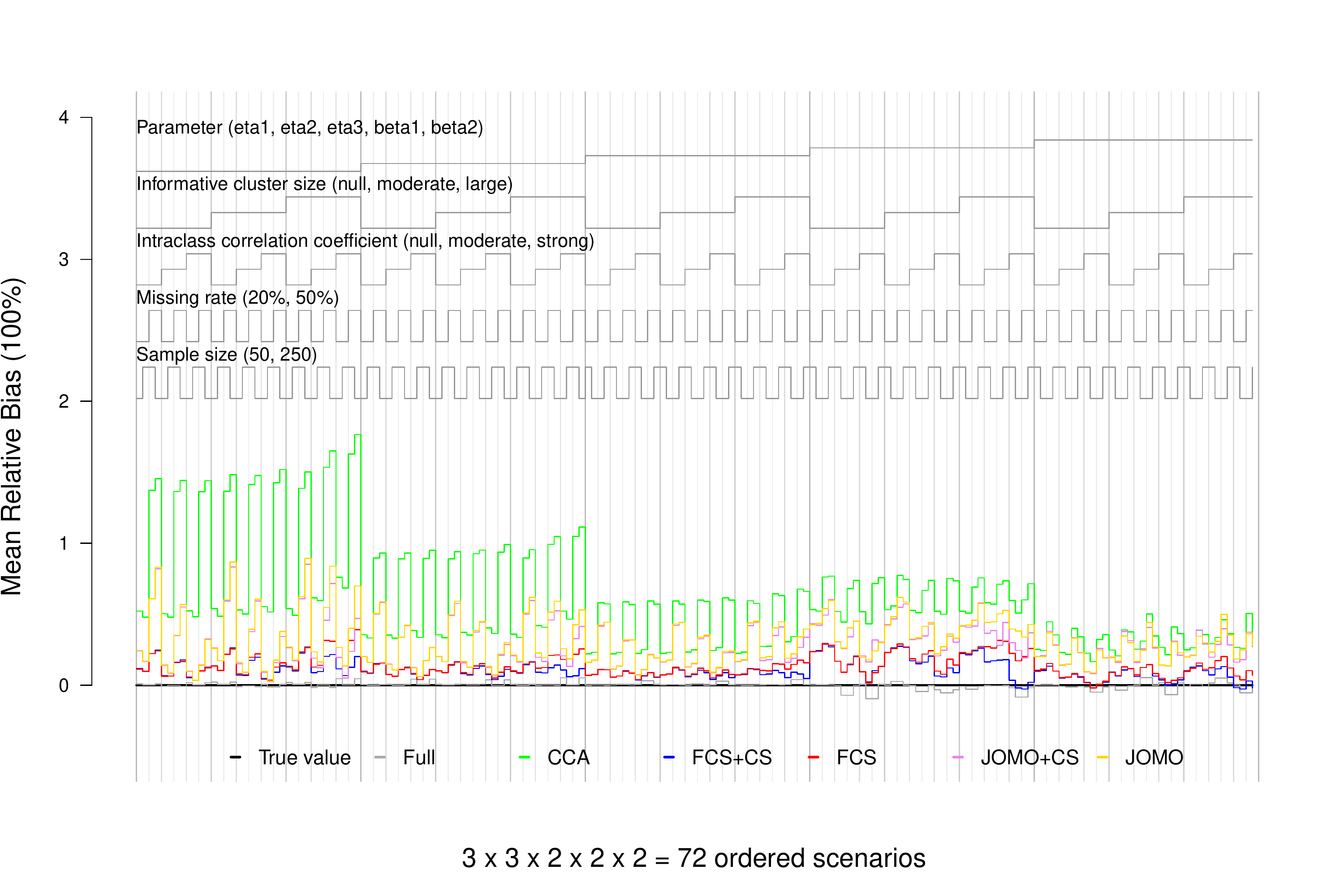}
   \caption{Mean relative bias of each imputation method and each parameter under different simulation scenarios when the missing data mechanism is MAR. The black line is the true value 0; grey line represents the results using the full data; grey line represents the results using complete case analysis; blue line represents the results using FCS; red line represents the results using FCS; purple line represents the results using JOMO+CS; orange line represents the results using JOMO.}
    \label{fig:relativebias}
\end{figure}

\begin{figure}[!htb]
    \centering
    \includegraphics[scale=0.6, trim={0 2cm 0 0},clip]{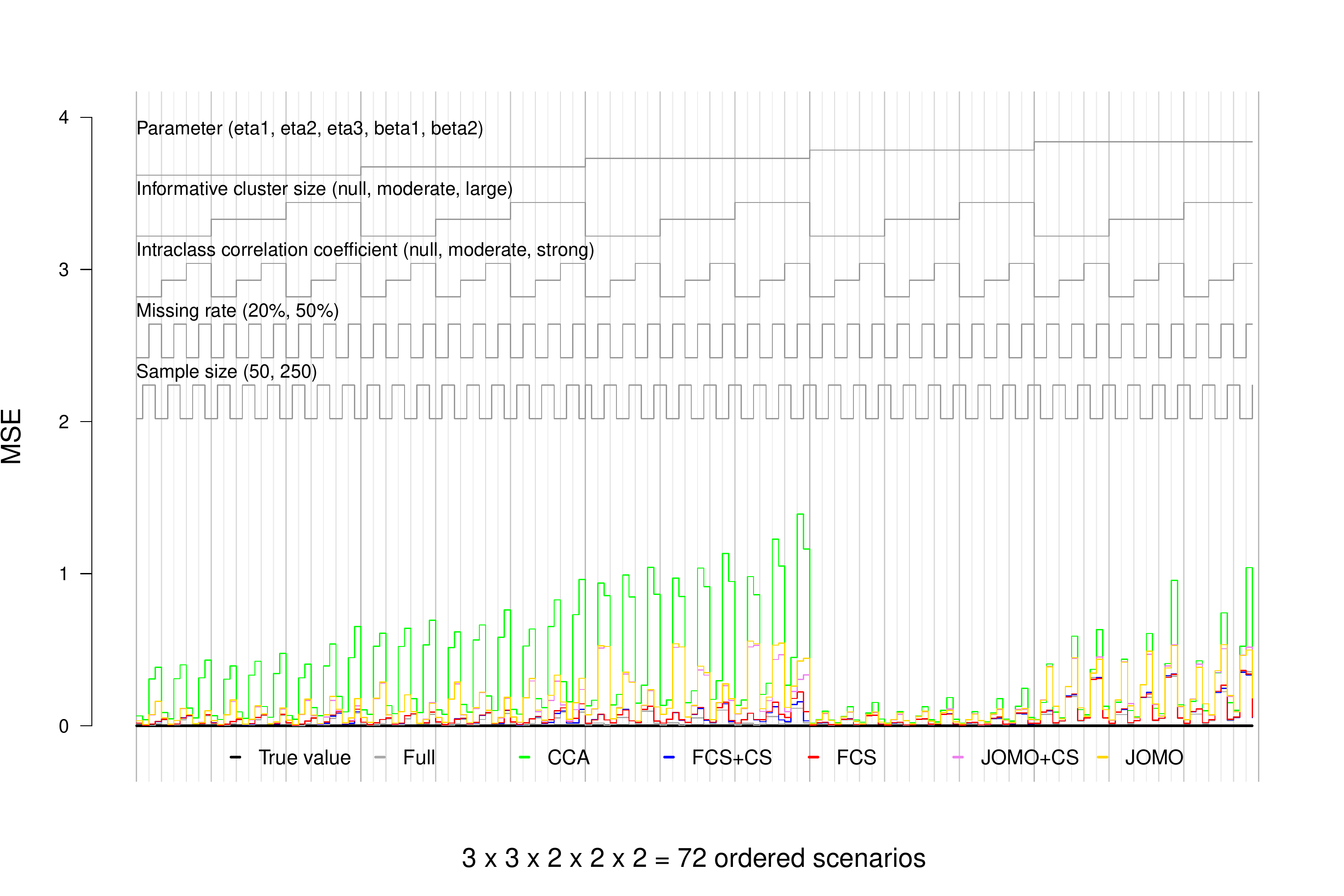}
 \caption{Mean squared error (MSE) of each imputation method and each parameter under different simulation scenarios when the missing mechanism is MAR. The black color is the true value 0; grey line represents the results using the full data; grey line represents the results using complete case analysis; blue line represents the results using FCS; red line represents the results using FCS; purple line represents the results using JOMO+CS; orange line represents the results using JOMO.}
    \label{fig:meanmse}
\end{figure}

Figure \ref{fig:relativebias} shows the mean relative biases of each imputation method and each parameter under different scenarios when the missing mechanism is MAR. CCA had the largest relative bias, followed by JOMO and JOMO+CS across all parameters. FCS+CS had the smallest mean relative bias in most scenarios, followed by FCS. The differences in mean relative bias between FCS+CS and FCS or between JOMO+CS and JOMO were negligible when the the degree of ICS was null or moderate. However, when the degree of ICS was large, FCS+CS had smaller mean relative bias than FCS, and JOMO+CS had smaller mean relative bias than FCS, indicating that including cluster size as a covariate in the imputation model for both FCS and JOMO improved the estimation accuracy. The effect of ICC on imputation was similar with ICS. The imputation methods were not affected by ICC when ICS was null. When ICS was moderate or large, CCA, JOMO and JOMO+CS performed worse as ICC increased, especially for the intercepts. It is noteworthy that this change was not very observable for FCS, given its good performance under most scenarios. The missing rate also had a large impact on mean relative biases. The mean relative biases for scenarios where the missing rate was 50\% were much larger than those where the missing rate was 20\%. The difference was more considerable for JOMO+CS, JOMO, and CCA, suggesting FCS was more reliable then JOMO when the missing rate was high. The mean relative bias was slightly lower when sample size was larger. Figure \ref{fig:meanmse} shows the MSE of each imputation method and each parameter under different scenarios when the missing mechanism is MAR. We can see that CCA produced the largest MSE over all scenarios, followed by JOMO. The performance of FCS and FCS+CS were similar to the results with full data, indicating the good performance of FCS. The MSEs of $\beta_1$ were close to 0 for all methods, which may be related to the scale of the true value. We also compared the empirical SEs and mean SEs of each method in Figures \ref{fig:empiricalse} and \ref{fig:meanse}. In general, the empirical SEs and mean SEs were similar. Both empirical SEs and mean SEs increased when the degree of ICS and ICC increased.

Simulation results of intercept $\eta_1$ and slope $\beta_1$ for scenario where ICC=0.3, ICS=0.1, missing rate= 20\%, and N=50 is shown in Table \ref{Tab:0.1_0.3_MAR}. 
The mean estimate of FCS+CS or FCS for the intercept $\eta_1$ was -0.37, which was closer to the true value compared to the JOMO+CS or JOMO. The mean relative biases were 7.06\% and 7.98\% for FCS and FCS+CS, respectively. The coverage probabilities of FCS and FCS+CS were both close to 95\%. Including cluster size in the imputation did not significantly improve the estimation accuracy, as the ICS was moderate. When looking into the details of more extreme cases when ICS was 0.4 and ICC was 0.6, we observed that including cluster size $n_i$ as a covariate in the imputation model shifted the mean $\hat\eta_1$ closer to the true value by 0.03 (Table \ref{Tab:0.4_0.6_MAR}). When the missing data mechanism was MNAR, all imputation methods had larger biases compared to when data were MAR (Table \ref{Tab:0.1_0.3_MNAR}). The mean estimates from CWGEE based on FCS and FCS+CS were -0.29, resulting in 26\% relative bias. The coverage probabilities of FCS and FCS+CS were around 93\%. The performance of FCS was still better than JOMO and CCA. When the missing data mechanism was MCAR, all imputation methods yielded unbiased estimates (Table \ref{Tab:0.1_0.3_MCAR}).

\begin{table}[!ht]
\centering
\resizebox{\linewidth}{!}{
\begin{tabular}{lllllllll}
  \hline
Parameter & Method & Mean Est & Mean SE & Empirical SE & Rel Bias (\%) & Cov Prob (\%) & MSE  \\ 
  \hline
  $\eta_1=-0.4$  & & & & & & & & \\
  \hline
&  Full &  -0.40 &   0.20 &   0.20 &  -0.97 &  95.50 &   0.04  \\ 
&   CCA &  -0.19 &   0.21 &   0.22 &  53.07 &  79.70 &   0.09  \\ 
&    FCS+CS &  -0.37 &   0.21 &   0.20 &  7.06 &  95.80 &   0.04  \\ 
&   FCS &  -0.37 &   0.21 &   0.20 &  7.98 &  95.90 &   0.04  \\ 
&  JOMO+CS & -0.34 &  0.23 &  0.22 & 15.09 & 94.18 &  0.05  \\ 
&   JOMO & -0.33 &  0.23 &  0.22 & 16.31 & 95.08 &  0.05  \\ 
   \hline  
  $\beta_1=-0.2$ & & & & & & & & \\
     \hline
&  Full &  -0.21 &   0.19 &   0.20 & -4.48 &  92.40 &   0.04  \\ 
 &  CCA &  -0.10 &   0.21 &   0.23 & 51.09 &  89.90 &   0.06  \\ 
&   FCS+CS &  -0.17 &   0.21 &   0.21 &  15.09 &  95.10 &   0.04  \\ 
& FCS &  -0.17 &   0.21 &   0.21 & 16.22 &  94.70 &   0.04  \\ 
&  JOMO+CS & -0.13 &  0.23 &  0.24 & 32.69 & 93.99 &  0.06  \\ 
&  JOMO & -0.13 &  0.24 &  0.24 & 33.50 & 93.57 &  0.06  \\ 
\hline
\end{tabular}
}
\caption{Results of intercept $\eta_1$ and slope $\beta_1$ when ICS=0.1, ICC=0.3, missing rate=20\%, N=50, MAR.}
\label{Tab:0.1_0.3_MAR}
\end{table}

\begin{table}[!ht]
\centering
\resizebox{\linewidth}{!}{
\begin{tabular}{lllllllll}
  \hline
Parameter & Method & Mean Est & Mean SE & Empirical SE & Rel Bias (\%) & Cov Prob (\%) & MSE \\ 
\hline
     $\eta_1=-0.4$ & & & & & & & & \\
  \hline
& Full &  -0.40 &   0.20 &   0.20 &  -0.97 &  95.50 &   0.04  \\ 
&   CCA & -0.10 &  0.22 &  0.24 & 74.61 & 71.21 &  0.15  \\ 
&   FCS+CS & -0.29 &  0.23 &  0.22 & 26.52 & 93.38 &  0.06  \\ 
&   FCS & -0.29 &  0.23 &  0.22 &  27.51 & 92.89 &  0.06 \\ 
&   JOMO+CS & -0.26 &  0.25 &  0.25 & 35.41 & 91.44 &  0.08  \\ 
&    JOMO & -0.26 &  0.25 &  0.25 & 36.02 & 91.40 &  0.09  \\ 
   \hline
        $\beta_1=-0.2$ & & & & & & & & \\
   \hline
& Full &  -0.21 &   0.19 &   0.20 &  -4.48 &  92.40 &   0.04  \\ 
&   CCA & -0.05 &   0.24 &   0.26 & 76.77 &  87.86 &   0.09  \\ 
&    FCS+CS & -0.13 &  0.24 &  0.22 & 36.58 & 96.09 &  0.05 \\ 
&  FCS & -0.12 &  0.24 &  0.22 & 40.06 & 95.40 &  0.05 \\ 
&  JOMO+CS &   -0.09 &   0.28 &   0.25 & 56.94 &  94.47 &   0.08  \\ 
&   JOMO & -0.08 &   0.28 &   0.26 & 59.03 &  94.03 &   0.08  \\ 
   \hline
\end{tabular}
}
\caption{Results of intercept $\eta_1$ and slope $\beta_1$ when ICS=0.1, ICC=0.3, missing rate=20\%, N=50, MNAR.} 
\label{Tab:0.1_0.3_MNAR}
\end{table}

\section{Real Data Application} \label{sec4}

Based on the analysis model in Equation (\ref{CAL}), we incorporated the level-1 variables described above, in additional with alveolar bone loss (ABL), mobility, and 
probing pocket depth (PPD). ABL, mobility and PPD are commonly used to quantify the severity of periodontitis, and are correlated with each other. Hence, ABL, mobility, and PPD were included as auxiliary variables to improve the imputation accuracy for CAL. In the imputation phase, we implement FCS+CS, FCS, JOMO+CS, JOMO and CCA. 

Table \ref{Tab:real_result} summarizes the results from the VADLS data. The effect of MetS based on the imputation method FCS+CS suggested that the odds of having a lower/healthier CAL score was approximately 20\% lower for patients free of MetS compared to patients with MetS, indicating that the presence of MetS is associated with a worse prognosis of periodontitis. Similar to the simulation studies, imputation with FCS+CS followed by the CWGEE model provided the smallest standard error. Recent studies have also reported the association between MetS and periodontal disease \cite{lamster2017periodontal}. However, such an association would not have been found if CCA was used onour data (OR=0.01, se=0.22).

\begin{table}[!ht]
\centering
\resizebox{\textwidth}{!}{\begin{tabular}{lllllllll}
  \hline
\textbf{Imputation} & \textbf{Intercept 1} & \textbf{Intercept 2} & \textbf{Intercept 3} & \textbf{age} & \textbf{smoking} & \textbf{Edu (level2)} & \textbf{Edu (level3)} & \textbf{MetS} \\ 
  \hline
  CCA & 2.68 (1.10) & 3.82 (1.10) & 4.81 (1.11) & -0.04 (0.01) & -0.26 (0.26) & -0.04 (0.26) & 0.4 (0.28) & 0.01 (0.22) \\ 
  FCS+CS & 2.68 (1.10) & 3.82 (1.11) & 4.79 (1.12) & -0.04 (0.01) & -0.33 (0.29) & 0.02 (0.24) & 0.47 (0.26) & -0.20 (0.20) \\ 
  FCS & 2.95 (1.11) & 4.09 (1.12) & 5.06 (1.13) & -0.04 (0.01) & -0.43 (0.27) & -0.06 (0.24) & 0.4 (0.26) & -0.14 (0.21) \\ 
  JOMO+CS & 2.45 (1.01) & 3.56 (1.02) & 4.52 (1.02) & -0.04 (0.01) & -0.26 (0.26) & -0.04 (0.24) & 0.42 (0.26) & -0.19 (0.2)  \\ 
  JOMO & 2.83 (1.18) & 3.96 (1.18) & 4.93 (1.19) & -0.04 (0.01) & -0.29 (0.27) & 0.07 (0.25) & 0.51 (0.26) & -0.11 (0.21) \\ 
   \hline
\end{tabular}}
\caption{The estimates (standard errors) for intercepts and covaraites in real data analysis.}
\label{Tab:real_result}
\end{table}

\section{Conclusions}\label{sec5}

In this study, we compared FCS, JM and CCA for imputing missing multilevel ordinal outcomes under different scenarios. Comprehensive simulation studies showed that FCS performed better than JM and CCA. FCS also provided a more reliable and stable performance with varying missing rates and ICC. We also saw that including cluster size as a covariate in the imputation model improved the estimation when the degree of ICS was large. MI methods are valid only when the missing data mechanism is MCAR and MAR. Nevertheless, both JM and FCS are better options for multilevel imputation than CCA.

Although both FCS and JM were based on Monte Carlo techniques, they are fundamentally different. FCS implemented by \textbf{blimp} imputes the ordinal data using a threshold-based latent probit approach. Since it imputes one variable at a time, it is more flexible and easier to adjust to different data types. On the other hand, the R package \textbf{jomo} uses a nominal probit model, even for imputing ordinal data, which tends to be less appropriate than FCS. Simulation studies conducted by Quartagno et al. showed that if the variables are truly ordinal, it gives good results with only a marginal loss in efficiency \cite{quartagno2019multiple}. However, we observed that the bias from jomo is not negligible compared to FCS when ICS exists. There is another software \textbf{REALCOM} that implements JM that could deal with incomplete multilevel ordinal outcomes, but unfortunately, it is only available for Windows users \cite{carpenter2011realcom}. 

This study has several limitations. First, we only compared the level-1 outcome missing in our data. Level-1 and level-2 covariates missing could make the imputation more complicated. Comparing the performance of JM and FCS when both covariates and outcomes are missing when ICS exists could be one of the future works. Second, although this study found that the bias introduced by JM using the R package \textbf{jomo} was not negligible, it is not fair to conclude that JM is worse than FCS when imputing multilevel ordinal outcomes. It is of interest to expand the package to deal with ordinal data. Last, our study focused on two-level clustered data, and did not assess the time effect typically observed in longitudinal studies. Wijesuriya et al. compared the methods for three-level data with time-varying cluster size for continuous exposures and outcomes \cite{wijesuriya2022multiple}. It would be of interest to extend the comparison to include ICS and ordinal outcomes. 

To conclude, this study compared JM and FCS for imputing multilevel ordinal outcomes when data is subject to ICS. We found that FCS is currently the optimal choice, and we should include the cluster size in the imputation model if there is potential for ICS. Our study provides as a further guide on choosing the imputation methods when imputing the multilevel ordinal outcomes.  

\section*{Acknowledgments}

We acknowledge Professor Raul Garcia who is the Principal Investigator and examiner for the Dental Longitudinal Study. The Dental Longitudinal Study and Normative Aging Study are components of the Massachusetts Veterans Epidemiology Research and Information Center which is supported by the VA Cooperative Studies Program. Views expressed in this paper are those of the authors and do not necessarily represent the views of the US Department of Veterans Affairs.

\subsection*{Financial disclosure}

This work was supported by the Dalla Lana School of Public Health Data Science Cluster and the Natural Sciences and Engineering Research Council of Canada - Discovery Grants Program (RGPIN-2022-05356).

\subsection*{Conflict of interest}

The authors declare no potential conflict of interests.

\bibliographystyle{unsrt}
\bibliography{sample}

\begin{thebibliography}{10}

\bibitem{seaman2014methods}
Shaun~R Seaman, Menelaos Pavlou, and Andrew~J Copas.
\newblock Methods for observed-cluster inference when cluster size is
  informative: a review and clarifications.
\newblock {\em Biometrics}, 70(2):449--456, 2014.

\bibitem{williamson2003marginal}
John~M Williamson, Somnath Datta, and Glen~A Satten.
\newblock Marginal analyses of clustered data when cluster size is informative.
\newblock {\em Biometrics}, 59(1):36--42, 2003.

\bibitem{benhin2005cwgee}
E.~Benhin, J.~N.~K. Rao, and A.~J. Scott.
\newblock Mean estimating equation approach to analysing cluster-correlated
  data with nonignorable cluster sizes.
\newblock {\em Biometrika}, 92(2):435--450, 2005.

\bibitem{mitani2019marginal}
Aya~A Mitani, Elizabeth~K Kaye, and Kerrie~P Nelson.
\newblock Marginal analysis of ordinal clustered longitudinal data with
  informative cluster size.
\newblock {\em Biometrics}, 75(3):938--949, 2019.

\bibitem{schafer1997analysis}
Joseph~L Schafer.
\newblock {\em Analysis of incomplete multivariate data}.
\newblock CRC press, 1997.

\bibitem{little2019statistical}
Roderick~JA Little and Donald~B Rubin.
\newblock {\em Statistical analysis with missing data}, volume 793.
\newblock John Wiley \& Sons, 2019.

\bibitem{rubin2004multiple}
Donald~B Rubin.
\newblock {\em Multiple imputation for nonresponse in surveys}, volume~81.
\newblock John Wiley \& Sons, 2004.

\bibitem{vanBuuren2018book}
Stef van Buuren.
\newblock {\em Flexible Imputation of Missing Data, Second Edition}.
\newblock Chapman and Hall/{CRC}, jul 2018.

\bibitem{novo2002analysis}
A~Novo and JL~Schafer.
\newblock Analysis of multivariate normal datasets with missing values.
\newblock {\em Ported to R by Alvaro A. Novo. Original by JL Schafer}, 2002.

\bibitem{harding6cat}
T~Harding, F~Tusell, and JL~Schafer.
\newblock cat: Analysis of categorical-variable datasets with missing values.
\newblock {\em URL http://CRAN. R-project. org/package= cat. R package version
  0.0-6.5.[p 205]}, 2012.

\bibitem{schafermix}
JL~Schafer.
\newblock mix: Estimation/multiple imputation for mixed categorical and
  continuous data.
\newblock {\em URL http://CRAN. R-project. org/package= mix. R package
  version}, pages 1--0, 2010.

\bibitem{pan}
Jing~Hua Zhao and Joseph~L. Schafer.
\newblock {\em pan: Multiple imputation for multivariate panel or clustered
  data}, 2018.
\newblock R package version 1.6.

\bibitem{quartagno2019jomo}
Matteo Quartagno, Simon Grund, and James Carpenter.
\newblock Jomo: a flexible package for two-level joint modelling multiple
  imputation.
\newblock {\em R Journal}, 9(1), 2019.

\bibitem{carpenter2011realcom}
James~R Carpenter, Harvey Goldstein, and Michael~G Kenward.
\newblock Realcom-impute software for multilevel multiple imputation with mixed
  response types.
\newblock {\em Journal of Statistical Software}, 45:1--14, 2011.

\bibitem{van2011mice}
Stef Van~Buuren and Karin Groothuis-Oudshoorn.
\newblock mice: Multivariate imputation by chained equations in r.
\newblock {\em Journal of statistical software}, 45:1--67, 2011.

\bibitem{audigier2019micemd}
Vincent Audigier and M~Resche Rigon.
\newblock micemd: a smart multiple imputation r package for missing multilevel
  data.
\newblock In {\em UseR! 2019}, 2019.

\bibitem{robitzsch2017package}
Alexander Robitzsch, Simon Grund, Thorsten Henke, and Maintainer~Alexander
  Robitzsch.
\newblock Package ‘miceadds’.
\newblock {\em R Package: Madison, WI, USA}, 2017.

\bibitem{enders2018fully}
Craig~K Enders, Brian~T Keller, and Roy Levy.
\newblock A fully conditional specification approach to multilevel imputation
  of categorical and continuous variables.
\newblock {\em Psychological methods}, 23(2):298, 2018.

\bibitem{enders2016multilevel}
Craig~K Enders, Stephen~A Mistler, and Brian~T Keller.
\newblock Multilevel multiple imputation: A review and evaluation of joint
  modeling and chained equations imputation.
\newblock {\em Psychological methods}, 21(2):222, 2016.

\bibitem{audigier2018multiple}
Vincent Audigier, Ian~R White, Shahab Jolani, Thomas~PA Debray, Matteo
  Quartagno, James Carpenter, Stef Van~Buuren, and Matthieu Resche-Rigon.
\newblock Multiple imputation for multilevel data with continuous and binary
  variables.
\newblock {\em Statistical Science}, 33(2):160--183, 2018.

\bibitem{wijesuriya2022multiple}
Rushani Wijesuriya, Margarita Moreno-Betancur, John Carlin, Anurika~Priyanjali
  De~Silva, and Katherine~Jane Lee.
\newblock Multiple imputation approaches for handling incomplete three-level
  data with time-varying cluster-memberships.
\newblock {\em Statistics in Medicine}, 2022.

\bibitem{kombo2017multiple}
Abdallah~Yusuf Kombo, H~Mwambi, and Geert Molenberghs.
\newblock Multiple imputation for ordinal longitudinal data with monotone
  missing data patterns.
\newblock {\em Journal of Applied Statistics}, 44(2):270--287, 2017.

\bibitem{kapur1972veterans}
Krishan~K Kapur, Robert~L Glass, Edward~R Loftus, John~E Alman, and Ralph~P
  Feller.
\newblock The veterans administration longitudinal study of oral health and
  disease: methodology and preliminary findings.
\newblock {\em Aging and Human Development}, 3(1):125--137, 1972.

\bibitem{kaye2016metabolic}
EK~Kaye, N~Chen, HJ~Cabral, P~Vokonas, and RI~Garcia.
\newblock Metabolic syndrome and periodontal disease progression in men.
\newblock {\em Journal of dental research}, 95(7):822--828, 2016.

\bibitem{gamonal2010clinical}
Jorge Gamonal, Carolina Mendoza, Iris Espinoza, Andrea Munoz, Ivan Urzua, Waldo
  Aranda, Paola Carvajal, and Oscar Arteaga.
\newblock Clinical attachment loss in chilean adult population: first chilean
  national dental examination survey.
\newblock {\em Journal of periodontology}, 81(10):1403--1410, 2010.

\bibitem{fitzmaurice2008longitudinal}
Garrett Fitzmaurice, Marie Davidian, Geert Verbeke, and Geert Molenberghs.
\newblock {\em Longitudinal data analysis}.
\newblock CRC press, 2008.

\bibitem{seaman2014review}
Shaun Seaman, Menelaos Pavlou, and Andrew Copas.
\newblock Review of methods for handling confounding by cluster and informative
  cluster size in clustered data.
\newblock {\em Statistics in medicine}, 33(30):5371--5387, 2014.

\bibitem{kenward1994ordinal}
M.~G. Kenward, E.~Lesaffre, and G.~Molenberghs.
\newblock An application of maximum likelihood and generalized estimating
  equations to the analysis of ordinal data from a longitudinal study with
  cases missing at random.
\newblock {\em Biometrics}, 50(4):945--953, 1994.

\bibitem{Williamson07}
J.~M. Williamson, H-Y. Kim, and L.~Warner.
\newblock Weighting condom use data to account for nonignorable cluster size.
\newblock {\em Annals of Epidemiology}, 17(18):603--607, 2007.

\bibitem{quartagno2019multiple}
Matteo Quartagno and James~R Carpenter.
\newblock Multiple imputation for discrete data: Evaluation of the joint latent
  normal model.
\newblock {\em Biometrical journal}, 61(4):1003--1019, 2019.

\bibitem{sterne2009multiple}
Jonathan~AC Sterne, Ian~R White, John~B Carlin, Michael Spratt, Patrick
  Royston, Michael~G Kenward, Angela~M Wood, and James~R Carpenter.
\newblock Multiple imputation for missing data in epidemiological and clinical
  research: potential and pitfalls.
\newblock {\em Bmj}, 338, 2009.

\bibitem{parzen2011generalized}
Michael Parzen, Souparno Ghosh, Stuart Lipsitz, Debajyoti Sinha, Garrett~M
  Fitzmaurice, Bani~K Mallick, and Joseph~G Ibrahim.
\newblock A generalized linear mixed model for longitudinal binary data with a
  marginal logit link function.
\newblock {\em The annals of applied statistics}, 5(1):449, 2011.

\bibitem{rubin1976inference}
Donald~B Rubin.
\newblock Inference and missing data.
\newblock {\em Biometrika}, 63(3):581--592, 1976.

\bibitem{lamster2017periodontal}
Ira~B Lamster and Michael Pagan.
\newblock Periodontal disease and the metabolic syndrome.
\newblock {\em International dental journal}, 67(2):67--77, 2017.

\end{thebibliography}

\clearpage

\section*{Supporting information}

\appendix

\begin{section}{Details on Simulating Multilevel Clustered Ordinal Data with Informative Cluster Size}
For every cluster $i$, 
\begin{enumerate}
    \item Sample $\boldsymbol{\omega}_i=(\omega_{i1}, \dots, \omega_{im})^{'}$ from a multivariate normal distribution, with mean vector $\boldsymbol{0}$ and variance matrix $\boldsymbol{\Sigma}$, where 
    $$\Sigma = \begin{pmatrix}
1 & \tau & \dots & \tau \\
\tau & 1 & \dots & \tau \\
\vdots & \vdots & \ddots & \vdots\\
\tau & \dots & \tau & 1 & 
\end{pmatrix}_{mxm}$$
$\tau$ is the correlation between each pair of units within a cluster. We used to exchangeable correlation structure to generate correlation between teeth. 
\item Compute $u_i = \phi(\boldsymbol{\omega}_i)$, where $\phi$ is the CDF of standard normal distribution.
\item Compute $b_i=\frac{1}{\phi} \log \frac{sin(\phi \pi u_i)}{sin(\phi \pi (1-u_i))}$, where $f_b(b_i | \phi) = \frac{1}{2\pi}\frac{sin(\phi \pi)}{cosh(\phi b_i)+ cos(\phi \pi)}$. $b_{ij}$ has marginal distribution and $b_{ij}$ and $b_{ik}$ are correlated due to the correlation imposed by $w_{ij}$. 
\item Compute the baseline level of risk $\lambda_i$ for each cluster such that $\lambda_i=\frac{\exp(\boldsymbol{\nu}_i \bar{b}_i))}{1+\exp(\boldsymbol{\nu}_i \bar{b}_i))}$, where $\bar{b}_i=\sum_j \frac{b_{ij}}{n_i}$.
\item Sample cluster size $n_i$ from a truncated binomial $(28,\lambda_i)$.
\item Generate the outcome $Y_ij$, which takes values from 1, 2, 3, and 4 from a multinomial distribution with a set of probability ($P_{ij1},P_{ij2}, P_{ij3}, P_{ij4}$) such that 

\begin{align*}
    P_{ij1} &= \Pr(Y_{ij}=1 | b_{ij}, X_{ij}, \boldsymbol{\beta})=\theta_1 \\
    P_{ij2} &= \Pr(Y_{ij}=2 | b_{ij}, X_{ij}, \boldsymbol{\beta})=\theta_2-\theta_1 \\
    P_{ij3} &= \Pr(Y_{ij}=3 | b_{ij}, X_{ij}, \boldsymbol{\beta})=\theta_3-\theta_2 \\
    P_{ij4} &= \Pr(Y_{ij}=4 | b_{ij}, X_{ij}, \boldsymbol{\beta})=1-\theta_3,
\end{align*}
where $\theta_c=\frac{\exp \{ b_{ij} + (\eta_c+X_{ij} \boldsymbol{\beta}) \phi^{-1} \} }{1+\exp \{ b_{ij} + (\eta_c+X_{ij} \boldsymbol{\beta}) \phi^{-1} \} }$, $c=1,2,3$.
\item Repeat for $i=1,\dots, N$ subjects.
\item Repeat the whole process for each auxiliary outcome with different values of $\eta_c$.
\end{enumerate}

\end{section}

\section{Supplementary Tables and Figures}
\setcounter{table}{0}
\renewcommand{\thetable}{S\arabic{table}}
\setcounter{figure}{0}
\renewcommand{\thefigure}{S\arabic{figure}}

\begin{table}[!ht]
    \centering
 \begin{tabular}{c|c|c|c|c}
   \hline
        Variables & Type & Categories & Summary Stats & Missing Rate \\
        \hline
        Age & subject-level & Median (range) & 76 (60, 98) & 0\% \\
        \hline
Smoking status & subject-level & Ever-smoker & 40 (17\%) & 0\% \\
\hline
\multirow{3}{4em}{Education} & \multirow{3}{5.3em}{subject-level} & High school & 62 (26\%) & \multirow{3}{1.3em}{0\%} \\
& & Some college & 86 (36\%)&  \\
& & College graduate & 93 (38\%) & \\
\hline
Metabolic Syndrome & subject-level & Yes & 95 (39\%) & 0\% \\
\hline
nteeth & subject-level & Median (range)  & 22 (1, 28) & 0\% \\
\hline
CAL & Tooth-level & levels & 4 & 19\% \\
\hline
PPD & Tooth-level & levels & 4 & 10\% \\
\hline
ABL & Tooth-level & levels & 6 & 25\% \\
\hline
Mobil & Tooth-level & levels & 4 & 0.2\% \\
\hline
    \end{tabular}
    \caption{Baseline Characteristics of covariates and outcome. PPD, ABL and Mobil are auxiliary variables used in the imputation phase.}
    \label{tab:char}
\end{table}

\begin{table}[!ht]
\centering
\resizebox{\textwidth}{!}{\begin{tabular}{lllllllll}
  \hline
Parameter & Method & Mean Est & Mean SE & Empirical SE & Rel Bias (\%) & Cov Prob (\%) & MSE \\ 
  \hline
    $\eta_1=-0.4$ & & & & & & & & \\
  \hline
& Full &  -0.38 &   0.30 &   0.29 &  4.58  &  95.20 &   0.08  \\ 
 & CCA & -0.10 &  0.29 &  0.32 & 76.03 & 76.68 &  0.19 \\ 
 & FCS+CS & -0.35 &  0.31 &  0.28 & 11.53 & 96.59 &  0.08 \\ 
 & FCS & -0.32 &  0.32 &  0.29 & 19.78 & 94.46 &  0.09\\ 
 & JOMO+CS & -0.32 &  0.34 &  0.30 & 19.62 & 95.61 &  0.10  \\ 
 & JOMO & -0.29 &  0.34 &  0.31 & 26.71 & 93.45 &  0.11 \\ 
   \hline
  $\beta_1=-0.2$ & & & & & & & & \\
   \hline
 & Full &  -0.22 &   0.25 &   0.29 & -8.30 &  90.70 &   0.08  \\ 
 & CCA &  -0.10 &   0.28 &   0.34 & 50.82 &  87.34 &   0.13  \\ 
 & FCS+CS &  -0.20 &   0.29 &   0.28 & -1.95 &  95.58 &   0.08  \\ 
 & FCS &  -0.17 &   0.30 &   0.29 & 17.16 &  95.87 &   0.08  \\ 
 & JOMO+CS &  -0.15 &   0.34 &   0.33 & 24.51 &  95.61 &   0.11 \\ 
 & JOMO & -0.12 &   0.34 &   0.31 & 38.12 &  96.12 &   0.10 \\ 
   \hline
\end{tabular}}
\caption{Results of intercept $\eta_1$ and slope $\beta_1$ when ICS=0.4, ICC=0.6, missing rate=20\%, N=50, MAR.}
\label{Tab:0.4_0.6_MAR}
\end{table}

\begin{table}[!ht]
\centering
\resizebox{\textwidth}{!}{\begin{tabular}{lllllllll}
  \hline
Parameter  & Method & Mean Est & Mean SE & Empirical SE & Rel Bias (\%) & Cov Prob (\%) & MSE \\
\hline 
$\eta_1=-0.4$ & & & & & & & & \\
\hline
& Full &  -0.40 &   0.20 &   0.20 &  -0.97 &  95.50 &   0.04  \\ 
 & CCA &  -0.40 &   0.21 &   0.21 &  -1.03 &  94.90 &   0.05  \\ 
 & FCS+CS &  -0.40 &   0.21 &   0.20 &  -0.05 &  95.30 &   0.04 \\ 
& FCS &  -0.40 &   0.21 &   0.20 &  0.48 &  95.50 &   0.04  \\ 
 & JOMO+CS &  -0.40 &   0.21 &   0.21 &  -0.74  &  95.80 &   0.04  \\ 
 & JOMO &  -0.40 &   0.21 &   0.21 &  -0.91 &  95.70 &   0.04  \\ 
  \hline
  $\beta_1=-0.2$ & & & & & & & & \\
  \hline
& Full &  -0.21 &   0.19 &   0.20 &  -4.48 &  92.40 &   0.04  \\ 
 & CCA &  -0.21 &   0.20 &   0.22 &  -6.43 &  92.30 &   0.05  \\ 
 & FCS+CS &  -0.21 &   0.19 &   0.20 &  -6.42 &  95.10 &   0.04  \\ 
& FCS &  -0.21 &   0.19 &   0.20 & -5.89 &  94.20 &   0.04  \\ 
 & JOMO+CS &  -0.21 &   0.20 &   0.21 & -5.19 &  94.00 &   0.04 \\ 
 & JOMO &  -0.21 &   0.20 &   0.21 &  -3.96 &  94.20 &   0.04  \\ 
   \hline
\end{tabular}}
\caption{Results of intercept $\eta_1$ and slope $\beta_1$ when ICS=0.1, ICC=0.3, missing rate=20\%, N=50, MCAR.} 
\label{Tab:0.1_0.3_MCAR}
\end{table}

\begin{figure}[!htb]
    \centering
    \includegraphics[scale=0.6, trim={0 2cm 0 0},clip]{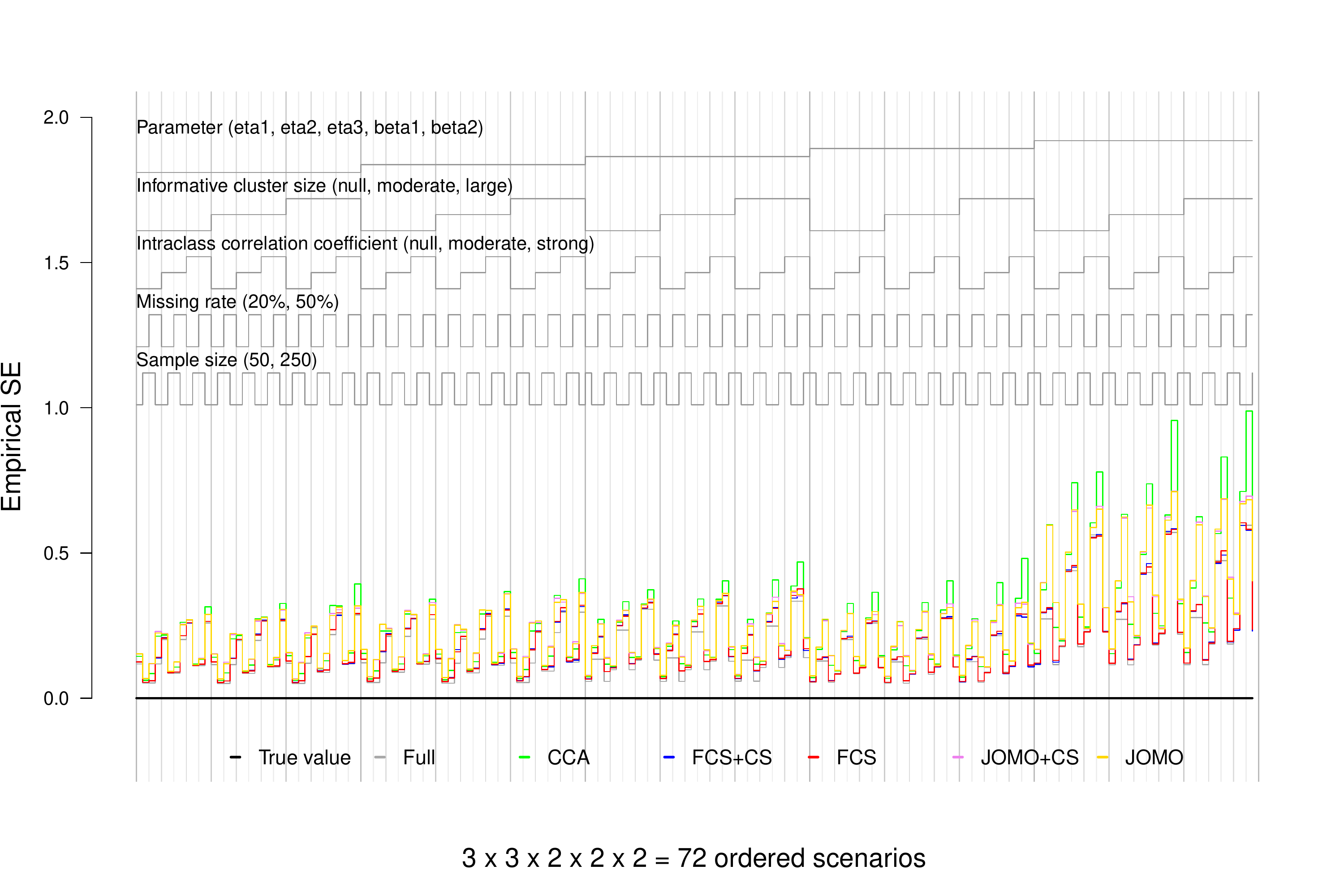}
    \caption{Empirical standard error of each imputation method and each parameter under different simulation scenarios when the missing mechanism is MAR. The grey lines represent the results using the full data; grey lines represent the results using complete case analysis; blue lines represent the results using FCS; red lines represent the results using FCS; purple lines represents the results using JOMO+CS; orange lines represent the results using JOMO.}
    \label{fig:empiricalse}
\end{figure}

\begin{figure}[!htb]
    \centering
    \includegraphics[scale=0.6, trim={0 2cm 0 0},clip]{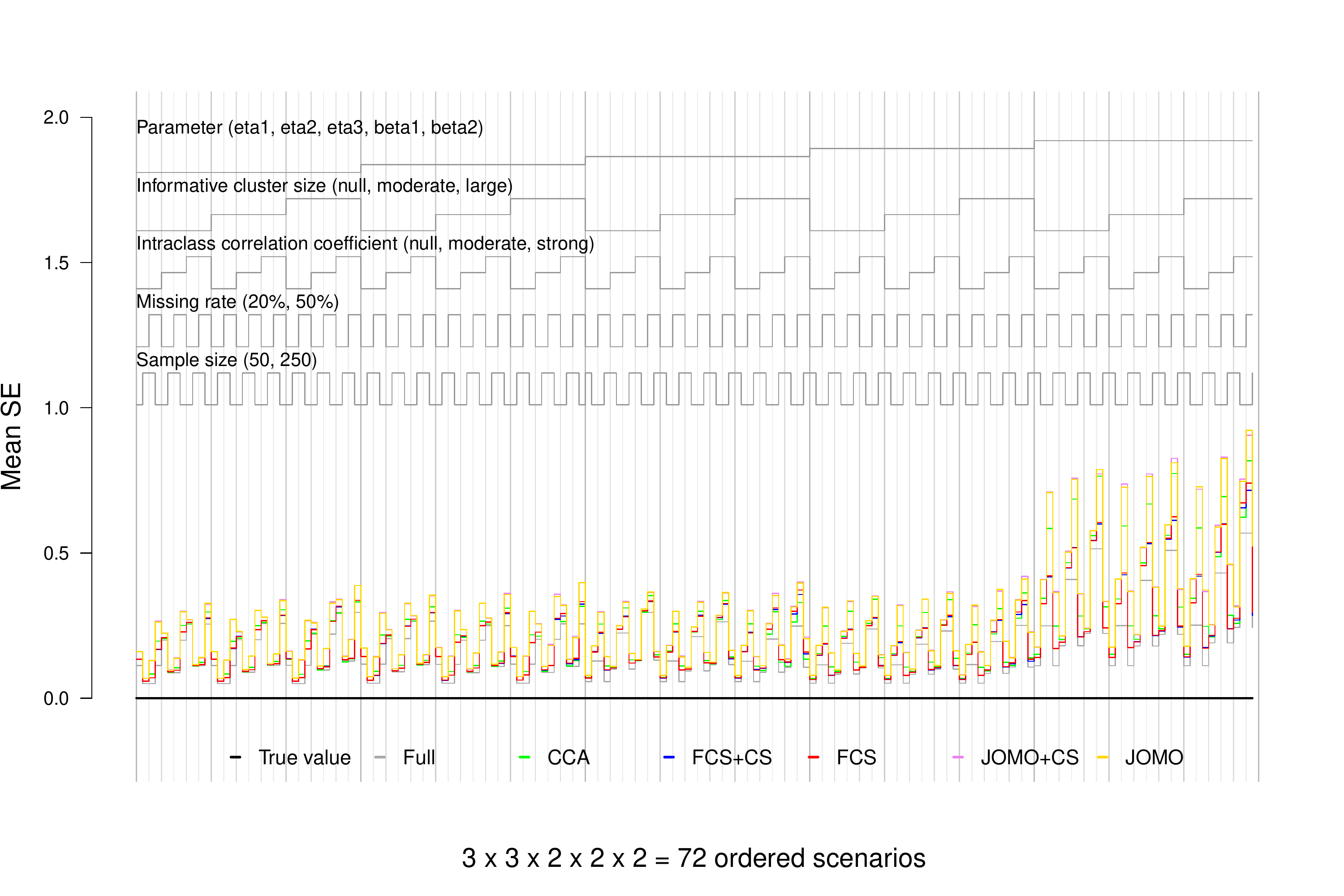}
    \caption{Mean standard error of each imputation method and each parameter under different simulation scenarios when the missing mechanism is MAR. The grey lines represent the results using the full data; grey lines represent the results using complete case analysis; blue lines represent the results using FCS; red lines represent the results using FCS; purple lines represents the results using JOMO+CS; orange lines represent the results using JOMO. }
    \label{fig:meanse}
\end{figure}

\end{document}